\documentclass{aastex}
\usepackage{mathrsfs}
\usepackage{amsfonts}
\usepackage{spr-astr-addons}
\usepackage{url}

\RequirePackage{color}

\newcommand{\emaila}{602061430@.com}

\begin{document}

\title{Jet/accretion and Unification for FSRQs/FRII and BL Lac objects/FRI radio galaxies}
\slugcomment{Not to appear in Nonlearned J., 45.}
\shorttitle{}
\shortauthors{Autors et al.}

\author{Yongyun Chen} \and \author{Xiong Zhang$^{\dag}$} \and \author{Haojing Zhang} \and \author{Dingrong
Xiong} \and\author{Xiaoling Yu} \and\author{Yongjuan Cha} \and\author{Bijun Li} \and\author{Xia Huang}\affil{Department of Physics, Yunnan Normal University, 650500,
\\Kunming, China\\$^{\dag}$e-mail:ynzx@yeah,net.}
\email{\emaila}

\begin{abstract}
We compiled a sample of 24 flat-spectrum radio quasars (FSRQs), 21 BL Lacs, 13 FRI and 12 FRII radio galaxies to study the jet power and broad-line luminosity relation, and the jet power and black hole mass relation. Furthermore, we obtain the histograms of key parameters. Our main results are as the follows:(i) We found that the FSRQs are roughly separated from BL Lacs by the Ledlow-Owen FRI/FRII dividing line in the $\rm{\log{P_{jet}}-\log{M}}$ plane. This result supports the unification model that BL Lacs unify with FRI radio galaxies and FSRQs unify with FRII radio galaxies.(ii)We found that the coefficient of the best-fit linear regression equation of $\rm{\log{L_{BLR}}-\log{P_{jet}}}$ relation is very close to 1 for our sample. The correlation between broad line luminosity and jet power is significant which supports that jet power has a close link with accretion.
\end{abstract}



\section{Introduction}
Based on the recent unified theory of active galactic nucleus (AGNs), one of the goals of the study of AGNs is to develop a unified scheme in which the different AGNs types might be understood in terms of variations in a few fundamental parameters, such as the black hole mass, the accretion rate, viewing angles (the angle between the jet and the line of sight), and the angular momentum per unit mass of the black hole (Meier 2002; Laor 2000; Boroson 2002; Barth et al. 2003). Blazars are characterized by a rapid optical variability, and a high and variable optical polarization (Xie et al. 1998, 1999; Zhang et al. 1998, 2004, 2007, 2008). Hence, the blazars are the most extreme class of AGNs. Blazars often divide into two subcategories of BL Lac objects (BL Lacs) and flat spectrum radio quasars (FSRQs). The classic division between FSRQs and BL Lacs is mainly based on the Equivalent Width (EW) of the emission lines. Objects with rest frame $(\rm{EW<5\AA})$ are classified as BL Lacs and $\rm{EW>5\AA}$ as FSRQs (see e.g. Urry \& Padovani 1995). Because of the many similarities of their characteristics, some authors have suggested that FSRQs and BL Lacs should be regarded as a single class that shows blazar behavior (e.g., Angel \& Stockman 1980). However, Browne (1989) thought that the FSRQs and BL Lacs should not be treated as a single category due to the different distribution of redshift and different emission-line strength. BL Lacs usually show weak or no emission lines (Burbidge \& Hewitt 1987). Conversely, FSRQs usually show strong and broad emission lines (Ledden \& O'Dell 1985).

The relation between FSRQs and BL Lacs has been studied by many authors (e.g., Sambruna et al. 1996; Ghisellini et al. 1998a,1998b; Xie et al. 2001a; Georganopoulos, Kirk, \& Masttichiadis 2001; D'Elia \& Cavaliere 2001; B$\ddot{o}$ttcher \& Dermer 2002). According to a study of the relation between radio quasars and a very small number of BL Lacs, Vagnetti, Cavaliere, \& Giallongo (1991) suggested that it may not be necessary to separate the FSRQs and BL Lacs. Padovani (1992) also studied the relation between FSRQs and BL Lacs. He found that the extend radio emissions and line luminosity are significantly different for FSRQs and BL Lacs. The intrinsic accretion rates are quite different among FSRQs, BL Lacs and radio galaxies. Ghisellini et al. (1998a,1998b) used a large sample of blazars to study these differences and found an increase of peak frequency from FSRQs to BL Lacs. D'Elia \& Cavaliere (2001) proposed that the BL Lacs are mainly powered by the rotational energy extraction of central super-massive black holes (SMBHs) via the Blandford-Znajek (BZ) mechanism, while FSRQs require a dominant contribution from the accretion of the central SMBHs. BL Lacs are supposed to be FRI type radio galaxies (Fanaro \& Riley 1974; Urry et al.1991). However, some evidences also shown that the parent population of BL Lacs should be a mix of FRI and FRII (Kollgaard et al. 1996), and the extended radio morphologies of BL Lacs were found to be of both FRI and FRII types (Kollgaard et al. 1992). Some emission lines from FRII are weak enough to be BL Lac-like (Laing 1994). Fan et al. (1997) found that BL Lacs, FRI and FRII have the same Hubble relation. Thus it may be appropriate to unify BL Lacs more generally with radio galaxies. It is well known that the broad-line luminosity is also an important parameter for understanding the nature of blazars besides the intrinsic bolometric luminosity. The broad-line luminosity can be taken as an indicator of the accretion power of the source (Celotti et al. 1997; Cao and Jiang 1999). Cao and Jiang (1999) found a significant correlation between radio and broad line luminosity for a sample of radio loud quasars. Czerny et al.(2004) analyzed the consequences of the hypothesis that the formation of broad line regions (BLRs) is intrinsically connected to the existence of cold accretion disks.

In this paper, through studying the jet power and black hole mass relation, the jet power and broad line luminosity relation, and comparing the key parameters of four classes, we get the relationship and physical distinction among FSRQs, BL Lacs, FRI and RII radio galaxies.

The paper is structured as follows: in Sect.2, we present the sample and data description; the results are in Sect.3; conclusions and discussions are in Sect.4.

\section{Sample and data description}
We compiled a group of blazars and radio galaxies with redshift and black hole mass. Firstly, we tried to get blazar sample with the minimum variability timescale. We mainly considered the sample of Xie et al. (2004a). They collected a group of blazars with the variable timescales from some of the literature. The relevant literature is listed in Table 1. Secondly, we considered the following sample of blazars to get the broad line data: Cao and Jiang (1999), Wang et al.(2002), Sbarrato et al.(2012), Maraschi and Tavechio (2003). Thirdly, We considered the sample of Zirbel \& Baum (1995) to get the emission line data for FR radio galaxies. Since radio galaxies almost have no broad emission line, Sbarrato et al. (2014) used $\rm{L_{nH_{\alpha}}/L_{bH_{\alpha}}}\leq10$ to derive the upper limits on broad emission luminosity for radio galaxies. Following Sbarrato et al.(2014), we obtain the upper limits of broad line luminosity for radio galaxies. We considered the following samples of radio galaxies to get black hole mass: Xie et al. (2004b), Woo \& Urry (2002). Finally, we got 45 blazars and 25 FR radio galaxies.

The jet power also can be derived from the lobe low frequency radio emission under the assumption of minimum energy arguments(e.g., Rawlings \& Saunders 1991; Willott et al. 1999). This approach now is widely used to estimate the
jet kinetic energy in AGNs. Meyer et al.(2011) used the following formula to estimate the cavity kinetic power,i.e.,
\begin{equation}
\rm{\log{P_{jet}}}=\rm{0.64(\pm0.09)(\log{L_{300}}-40)+43.54(\pm0.12)}
\end{equation}
where $\rm{L_{300}}$ is the extend luminosity at 300 MHz, the unit of jet power is erg $\rm{s^{-1}}$. Meyer et al. (2011) defined a core dominance parameter with $\rm{R_{CE}}=\rm{\log{(L_{core}/L_{ext})}}$ at 1.4 GHz. As shown in Figure 5 of Meyer et al. (2011), the typical $\rm{R_{CE}}$ for blazars is 0.5, i.e., $\rm{L_{core}/L_{ext}=3}$, and for radio galaxies is -1.5,i.e.,$\rm{L_{core}/L_{ext}}=0.03$. Zhang et al. (2014) used 5 GHz core luminosity to derived the 1.4 GHz core luminosity with the spectral index of $\rm{\alpha=0.5}$, and then they employ the core dominance parameter
$\rm{R_{CE}}$ to obtain the extended luminosity at 1.4 GHz. They finally obtain $\rm{L_{300} MHz}$ by
using a radio spectral index of $\rm{\alpha=1.2}$ and calculated the jet kinetic power by equation (1). Following Zhang et al. (2014), we use a 5 GHz core luminosity to calculate the jet kinetic power of our sample by using the equation (1). The relevant references of 5 GHz core luminosity are listed in Table 1 and 2.

We also noticed that our sample mainly entails the lower redshift objects. The redshift range is from 0 to 2. The main range is $\rm{0<z<1}$. There are no high redshift objects in our sample. Clearly, there is an observation bias in the distribution of redshift. The band for the observed minimum variation timescales is mainly the optical band; this may lead to the missing of the high redshift objects. Shaw et al. (2010) suggested that the BL Lacs have no measured redshift due to the lack of any detectable feature in their optical spectrum, despite the use of 10 m class optical telescopes for the spectroscopy identification campaign. Blandford \& Rees (1978) had originally suggested that because of blazars have a very bright, Doppler-boosted synchrotron continuum, the redshift of blazars is more difficult to be determine compared with those of normal AGN.
\subsection{The Sample of blazars}
An extensive literature search for reported balzars was performed. We searched for the shortest timescale of variation reported and selected blazars based on two criteria. Firstly, we searched for possible quasi-periodic variability in the observed flux curves. We defined the quantity $\rm{P_{min}}$ to be the minimum time interval between two local maxima in the flux curves required for a flux change of $\rm{{\Delta{I}}/I}\geq50\%$ between local maximum and minimum. In addition, the amplitude of variability on this timescale must be larger than $5\sigma$, where $\sigma$ is the maximum total observational rms error (Xie et al. 2001b). Secondly, if the observed light curve was found to have only one maximum and one minimum, we defined twice the time interval between this maximum and minimum as the minimum timescale for flux variations (Xie et al. 2004a).

Celotti et al. (1997) considered the sample of radio loud AGNs, which includes 105 sources with VLBI angular diameter data available in the literature. The broad-line luminosity given in Celotti,Padovani \& Ghisellini (1997) were derived by scaling several strong emission lines to the quasar template spectrum of Francis et al. (1991), and used $\rm{Ly\alpha}$ as a reference. Francis et al. (1991) set $\rm{Ly\alpha}$ flux contribution to 100, the relative weight of $\rm{H\alpha}$, $\rm{H\beta}$, MgII, and CIV lines respectively, to 77, 22, 34, and 63. The total broad line flux was fixed at 555.76. When more than one line was presented, they calculated the simple average of broad-line luminosity estimated from each line. The rest of authors in our sample also adopted the method proposed by Celotti,Padovani \& Ghisellini (1997) and similar processes to obtain broad line luminosity. Based on the flux integration and SED fitting, Woo \& Urry (2002) calculated the the bolometric luminosity of AGNs. Xie et al. (1991a,1991b)and Xie et al. (2004a) calculated the Doppler factor by using the variation of observed bolometric luminosity $(\rm{\Delta{L_{obs}}})$, the corresponding minimum variability timescale and the spectra index. Fabian \& Rees (1979) argued that a luminosity change $\rm{\Delta{L_{obs}}}$ on a timescale $\rm{\Delta{t_{min}}}$ should satisfy the limit $\rm{\eta\geq5.0\times10^{-43}\Delta{L}/\Delta{t_{min}}}$, where $\eta$ is the efficiency of the conversion of accreted matter into energy for a spherical, homogeneous, beam region. Assuming the super-massive black hole is a Kerr black hole with the maximum possible speed of rotation for the accretion model, the black hole mass in column (10) of table (1) was estimated by the following formula:
\begin{equation}
{\rm{M_{BH}=1.62\times10^{4}\frac{\delta}{1+z}\Delta t_{min}^{obs}M_{\odot}}}
\end{equation}
presented in the paper of Xie et al.(2002). The definition of $\rm{\Delta t_{min}^{obs}}$ has been given in detail by Elliot \& Shapiro (1974).

In this paper, the cosmological parameters ${\rm{H_{0}}}=70{\rm{kms^{-1}Mpc}}^{-1}$,  $\rm{\Omega_{m}=0.3}$, and $\rm{\Omega_{\Lambda}=0.7}$ have been adopted. The relevant data are listed in Table 1: column (1) Name: name of source; column (2) Class: gives the class (flat-spectrum radio quasars=FSRQ, BL Lac objects=BL); column (3) z: redshift; column (4) $\rm{\log{\Delta t_{\rm{min}}^{\rm{obs}}}}$: the minimum variability timescale in units of seconds; column (5) Ref: the references of minimum variability timescale; column (6) ${\rm{\log L_{BLR}}}$: the broad-line luminosity in units of $\rm{erg s^{-1}}$; column (7) Ref: the references of broad-line luminosity; column (8) ${\rm{\log L_{core}^{5GHz}}}$: the core luminosity at 5 GHz in units of $\rm{erg s^{-1}}$;  column (9) Ref: the references of 5 GHz core luminosity; column (10) ${\rm{\log (M/M_{\odot})}}$: the black hole mass; column(11) $\rm{\log{M_{\sigma}/M_{\odot}}}$: the black hole mass from stellar velocity dispersion.
\subsection{The Sample of Radio Galaxies}
Because the FRI and FRII radio galaxies do not have a rapid optical variability or unbeamed characteristics, the black hole mass can not be estimated by the observed variability timescale (Xie et al. 2002b). The black hole mass of radio galaxies in our sample have been derived from observed stellar velocity dispersions by Woo \& Urry (2002). Woo \& Urry (2002) used the following formula to calculate the black hole mass
\begin{equation}
{\rm{M_{BH}=1.349\times10^8M_{\odot}(\sigma/200kms^{-1})^{4.02}}}.
\end{equation}
The catalog of emission lines of radio sources was compiled by Zirbel \& Baum (1995). The relevant data are listed in Table 2: column (1) Name: name of the sources; column (2) Class: gives the class of sources; column (3) z: redshift; column (4) ${\rm{\log (M/M_{\odot})}}$: The black hole mass; column (5) Ref: references of the black hole mass data; column (6) ${\rm{\log L_{BLR}}}$: the broad-line luminosity in units of $\rm{erg s^{-1}}$; column (7) Ref: references of the broad-line luminosity data; column (8) ${\rm{\log L_{core}^{5GHz}}}$: the core luminosity at 5GHz in units of $\rm{erg s^{-1}}$; column (9) Ref: the references of 5 GHz core luminosity.

\section{Results}
\subsection{The distributions}
The redshift distributions of the various classes are shown in Fig.1. The mean redshifts are z=0.75$\pm$0.51 for FSRQs, z=0.26$\pm$0.24 for BL Lacs, z=0.05$\pm$0.03 for FRI ,and z=0.06$\pm$0.04 for FRII radio galaxies. The FSRQs have much higher average redshifts than BL Lacs. The FRI have lower average redshifts than FRII radio galaxies.

The black hole mass is a fundamental property of AGNs. Figure 2 shows the distributions of the black hole mass of various classes. The mean black hole masses are $\rm{10^{8.71\pm0.52}M_{\odot}}$ for FSRQs, $\rm{10^{8.01\pm0.39}M_{\odot}}$ for BL Lacs, $\rm{10^{8.50\pm0.27}M_{\odot}}$ for FRI, and $\rm{10^{8.29\pm0.39}M_{\odot}}$ for FRII radio galaxies. Compared with BL Lacs, FSRQs have much higher average black hole mass. The FRII have lower average black hole mass than FRI radio galaxies. These results suggest that the different observational features of FSRQs and BL Lacs may be mainly dominated by different black hole mass. Ghisellini and Tavecchio (2008) got that the average black hole masses are $\rm{10^{8.74\pm0.49}M_{\odot}}$ for FSRQs, $\rm{10^{8.57\pm0.43}M_{\odot}}$ for BL Lacs, $\rm{10^{8.82\pm0.50}M_{\odot}}$ for FRI and $\rm{10^{8.57\pm0.46}M_{\odot}}$ for FRII radio galaxies. We find that their results have a little difference with us. This difference may be explained by the different sample. Since FRI have larger average black hole masses than FRII, they might be older or have accreted at a greater rate in the past (through e.g. merging), and therefore it is conceivable to argue that at least a fraction of them were FR II radio galaxies in the past ( Urry \& Padovani 1995). The black hole mass play an important role in the formation of jet for AGN. Foschini (2014) studied the relationship between jet and black hole mass and got that different black hole mass may lead to different jet model.

Broad-line luminosity is another important parameters for AGNs. The broad-line luminosity distributions are given in Fig.3 for various classes. The average broad-line luminosities are $\rm{10^{45.09\pm0.44}erg s^{-1}}$ for FSRQs, $\rm{10^{42.70\pm1.22}erg s^{-1}}$ for BL Lacs, $\rm{10^{42.01\pm0.80}erg s^{-1}}$ for FRI, and $\rm{10^{42.11\pm0.97}erg s^{-1}}$ for FRII radio galaxies. Compared with FSRQs, BL Lacs have much lower average broad line luminosity. The FRII have higher average broad line luminosity than FRI radio galaxies.

The accretion rate is an important parameter for understanding the nature of AGNs. Based on accretion disk theory and the relativistic beaming model of AGNs, we know that the radiation efficiency is associated with Eddington accretion. The depletion of accretion power can lead to a decrease of the Eddington ratio, which may result in the transition of a different accretion model (Wu et al. 2002). Figure 4 shows the distributions of the Eddington ratios for various classes (${\rm{L_{bol}/L_{Edd}}}$, ${\rm{L_{Edd}}}={\rm{\frac{-4{\pi}GMm_{p}c}{{\sigma}T}}}={\rm{1.25\times10^{38}(\frac{M}{M_{\odot}})ergs^{-1}}}$, $\rm{L_{bol}\approx10L_{BLR}}$ from Netzer (1990)). From Fig.4, we can find that the Eddington ratios of the various classes are different. The average Eddington ratios are $\rm{10^{-0.85\pm0.50}}$ for FSRQs, $\rm{10^{-2.36\pm0.76}}$ for BL Lacs, $\rm{10^{-3.60\pm0.98}}$ for FRI , and $\rm{10^{-3.29\pm1.04}}$ for FRII radio galaxies. Compared with BL Lacs, FSRQs have higher average Eddington ratios. The FRII have higher average Eddington ratios than FRI radio galaxies. Xu et al. (2009) also found that the radio quasars have higher Eddington ratios than BL Lacs. Ghisellini and Celotti (2001) suggested that the FRII have higher accretion rate than FRI. A possible interpretation of this phenomenon is the exhaustion of the gas stockpile available for accretion in the host galaxy (Cattaneo et al. 1999; Haehnelt and Kauffmann 2000). Cattaneo et al. (1999) and Haehnelt and Kauffmann (2000) argued that with the gradual depletion of the gas, FSRQs (mainly fueled by accretion) will switch to BL Lacs (mostly fed by Kerr black hole rotational energy extraction via the BZ mechanism). There is a dense environment around the FSRQs and a rapid black hole growth, which contain a significant amount of gases in their central regions. These gases, as well as producing the observed emission lines, are used to produce the violent radio and optical activity. When these gases are used up or expelled, the emission lines will become very weak or disappear.

The jet power distributions of various classes are shown in Fig.5. The average jet powers are $\rm{10^{45.60\pm0.77}}$ for FSRQs, $\rm{10^{44.01\pm0.87}}$ for BL Lacs, $\rm{10^{43.87\pm0.61}}$ for FRI and $\rm{10^{43.63\pm0.51}}$ for FRII radio galaxies. Compared with BL Lacs, FSRQs have higher average jet powers. The FRII have lower average jet powers than FRI radio galaxies.
\subsection{Relation between jet power and black hole mass}
Ledlow and Owen (1996) got the division between FRI and FRII radio galaxies in the plane of total radio luminosity and optical luminosity of the host galaxy. Xu et al.(2009) got this dividing line
\begin{equation}
\rm{\log{Q_{jet}}(erg s^{-1})}=\rm{1.13M_{bh}+33.18+1.5\log{f}}
\end{equation}
in the $\rm{M_{bh}-Q_{jet}}$ plane (see Wu \& Cao 2008 for details). In Figure 6, we plot the relation between the black hole mass and jet power for FSRQs and BL Lacs with f=20. It is found that FSRQs can be roughly separated from BL Lacs by the FRI/FRII dividing line. This result supports the BL Lacs/FRI and FSRQs/FRII unification schemes. Ghisellini \& Celotti (2001) found that the FRI and FRII can be divided in the radio-host galaxy optical luminosity plane. Xu et al. (2009) found that the BL Lacs can be roughly separated from radio quasars by the FRI/FRII dividing line.
\subsection{Relation between jet power and broad line luminosity}
Figure 7 shows broad line luminosity as a function of jet power. Linear regression is applied to the relevant data to analyze the correlation between broad line luminosity and jet power with $95\%$ confidence level:
\begin{equation}
\rm{\log{L_{BLR}}}=\rm{1.10(\pm0.10)\log{P_{jet}}-5.99(\pm4.85)}
\end{equation}
The results show a strong correlation between broad line luminosity and jet power(r=0.80,P$<$0.0001,N=55). The result of a Pearson partial analysis also shows that there is still a significant correlation between broad line luminosity and jet power (N=55, P$<$0.0001, r=0.528). The result suggests a close link between the formation of jets and accretion.
\section{Conclusions and discussions}
According to the recent unified theory of AGN, one of the goals of the study of AGNs is to develop a unified scheme in which the different AGNs types might be understood in terms of variations in few fundamental parameters, such as the black hole mass, Eddington ratios and viewing angles (Meier 2002). In this work, we try to find the relationship between FSRQs/FRII and BL Lacs/FRI radio galaxies. Our main results are as follows:

(i) Xie et al.(2005) collected a sample of 21 AGNs, for which the minimum variability timescales have been obtained from some literatures and their black hole mass $(\rm{M_{\sigma}})$ have been well estimated from the stellar velocity dispersion by Woo \& Urry (2002). According to a comparison these two methods, Xie et al.(2005) found that using the Kerr black hole theory leads to a small difference between $\rm{M_{k}}$ and $\rm{M_{\sigma}}$, not exceeding one order of magnitude. On the basis of the work of Xie et al.(2005), we performed a comparison of the black hole mass estimated by the method of Xie et al.(2002) and the method of Woo \& Urry (2002). Based on the paper of Woo \& Urry (2002), we found 20 blazars that have relevant black bole masses. Data from this comparison are listed in column (14) of Table 1. From Table 1, we can see that the difference between M and $\rm{M_{\sigma}}$ is very small, the largest difference of them is less than one order of magnitude. The range of black hole mass in our sample is from $10^{7.25}$ to $10^{9.50}{\rm{M_{\odot}}}$. The FSRQs have much higher average black hole mass than BL Lacs. The FRI have higher average black hole mass than FRII radio galaxies. Many authors have studied the black hole mass and suggested that the black hole mass is very important for understanding the emission of AGNs (Cao \& Jiang 2002; Fan et al.2009; Fan et al. 2009; Wang et al. 2006; Wu et al. 2002 ). Foschini (2014) studied the relationship between jet and black hole mass. He suggested that different black hole mass may lead to different jet model.

(ii) In this paper, we only considered the sources of known redshift. As discussed in Ghisellini et al. (2011), if the BL Lacs unknown redshift will turn out to be at ${\rm{z\sim<0.5-1}}$, and it would lie in the region already occupied by the other BL Lacs which have measured redshifts (see Fig.1 and Fig.2
of Ghisellini et al. (2011)). The unknown redshift BL Lacs would fit in the phenomenological blazar sequence, they
would be with low and moderate luminosity, with the majority having a flat ${\rm{\gamma-ray}}$ slope. The fact that their spectrum is featureless implies that their emission lines are intrinsically weak. Instead, if ${\rm{z\sim2}}$, these unknown redshift sources would have a ${\rm{\gamma-ray}}$ luminosity as large as the powerful FSRQs. Ghisellini et al. (2011) suggested that there are two possibility:(i) They have powerful lines and an even more powerful continuum, swamping the emission lines. In this case, and for average black hole masses, their ${\rm{L_{BLR}/L_{Edd}}}$ is large, as it is ${\rm{L_{\gamma}/L_{Edd}}}$. They together with the other FSRQs (these objects in the ``FSRQs quadrant''); (ii) They have intrinsically weak lines, and average black hole masses. In this case ${\rm{L_{BLR}/L_{Edd}}}$ would be small, but ${\rm{L_{\gamma}/L_{Edd}}}$ is large. They would then occupy the bottom right part in the ${\rm{L_{BLR}/L_{Edd}}}$-${\rm{L_{\gamma}/L_{Edd}}}$ plane (see Fig.3 of Ghisellini et al. (2011)). In this case there would be no correlation between ${\rm{L_{BLR}/L_{Edd}}}$ and ${\rm{L_{\gamma}/L_{Edd}}}$, now devoid of sources. These sources would have a powerful non-thermal emission, possibly extremely beamed, even if their accretion disk is weak. Ghisellini et al. (2011) suggested a new classification criterion distinguishing BL Lacs from FSRQs based on the luminosity of the broad emission lines measured in Eddington units. This new classification scheme is not affected by the possible redshift values of the BL Lacs with unknown redshift. On the contrary, it can strengthen possible problems concerning the blazar sequence. We studied the distributions of the redshift, broad-line luminosity, Eddington ratios and jet power.

(iii) In additional, we also found that the FSRQs have higher accretion, broad line luminosity and jet power than BL Lacs. A reasonable explanation for our results is as follows. FSRQs occur in the earlier phase. They have a powerful disk and jet, and high accretion. With time, the FSRQs will have lower accretion rate, a less efficient disk, and shrinking BLR. It is possible that some transitions between FSRQs and BL Lacs appear with moderate BLR luminosity. When the accretion rate decreases below the critical value (i,e., $\rm{L_{bol}/L_{Edd}\sim0.01}$), the accretion changes mode, becoming radiative inefficient and FSRQs become BL Lacs. BL Lacs have a weak disk and weaker emission line emitted closer to the black hole. Dissipation in the jet occurs outside the BLR (if it exists at all). In other words, the quasars evolves into BL Lacs(Cavaliere \& D'Elia 2002). Cao (2002) got that the the evolutionary sequence of blazars is from a flat-spectrum radio quasars to a low-energy-peaked BL Lac object to a high-energy-peaked BL Lac object.

(iv) It is well know that the broad-line luminosity can be taken as an indicator of the accretion power of the source (Celotti et al. 1997). From our results, we can see that the correlation between broad line luminosity and jet power is significant, which supports that jet power has a close link with accretion. According to Ghisellini (2006), if relativistic jet are powered by a poynting flux, under some reasonable assumptions, the BZ jet power can be written
\begin{equation}
\rm{L_{BZ,jet}}\sim\rm{(\frac{\alpha}{m})^{2}\frac{R_{s}^{3}}{HR^{2}}\frac{\varepsilon_{ B}}{\eta}\frac{L_{disk}}{\beta_{r}}}
\end{equation}
where $\rm{\frac{\alpha}{m}}$ is the specific BH angular momentum: $\rm{R_{S}}=\rm{\frac{2GM_{BH}}{c^{2}}}$ is the Schwarzschild radius; H is the disk thickness; R is the radius; $\varepsilon_{B}$ is the fraction of the available gravitational energy; $\eta$ is the accretion efficiency; $\rm{L_{disk}}$ is the observed luminosity of accretion disk; $\rm{\beta_{r}}$ is the radial infalling velocity. The maximum BZ jet power can then be written as (Ghisellini 2006)
\begin{equation}
\rm{L_{jet}}\sim\rm{\frac{L_{disk}}{\eta}}.
\end{equation}
In additional, on the basis of current theories of accretion disks, the BLR is ionized by the radiation of the accretion disk. We have
\begin{equation}
\rm{L_{disk}\approx10L_{BLR}}.
\end{equation}
From Eqs. (7) and (8), we have
\begin{equation}
\rm{L_{BLR}\sim0.1\eta L_{jet}}
\end{equation}
From Eq.(9), we have
\begin{equation}
\rm{\log{L_{BLR}}}=\rm{\log{L_{jet}}+\log{\eta}+const}.
\end{equation}
Equation (10) shows that the theoretical predicted coefficient of the $\rm{\log{L_{BLR}}-\log{L_{jet}}}$ relation is 1. Using linear regression analysis, we obtain $\rm{L_{BLR}}\sim\rm{(1.10\pm0.10)\log{P_{jet}}}$ for our sample, which is very close to 1. Since $\eta$ is not the sample for all objects, $\rm{\log{\eta}}$ may contribute to the dispersion around the linear relation (Xie et al. 2007). Our results are also consistent with the results of D'Elia \& Cavaliere (2001), Xie et al. (2003), Cavaliere \& D'Elia (2002), and B$\ddot{o}$ttcher \& Dermer (2002). Rawlings \& Saunders (1991) found a close relationship between total kinetic power and narrow line luminosity for FRII and FRI radio galaxy. Sbarrato et al. (2014) also found correlation between broad line luminosity and radio luminosity for Fermi blazars and radio galaxies.

(v) Browne (1989) proposed that BL Lacs and OVV quasars are distributed differently; thus, their parent populations must be different. Two unified schemes have been proposed. One is the unification of BL Lacs and FRI radio galaxies, the other is the unification of OVV quasars and FRII radio galaxies (Browne 1989). Much evidence supporting the unified model has been presented by Xie et al. (1993,1994), Urry et al. (1991), and Owen et al.(1996). The results of Fig.s (6) provide evidence for the unified models of FSRQs/FRII and BL Lacs/FRI.

\acknowledgments We thank the anonymous referees for valuable
comments and suggestions. We are very grateful to the Science
Foundation of Yunnan Province of China(2012FB140,2010CD046). This work is supported by
the National Nature Science Foundation of China (11063004,11163007,U1231203), and the High-Energy Astrophysics Science and Technology Innovation Team of Yunnan Higher School and Yunnan Gravitation Theory Innovation Team (2011c1). This research has made use of the NASA/IPAC Extragalactic Database (NED), that is operated by Jet Propulsion Laboratory, California Institute of Technology, under contract with the National Aeronautics and
Space Administration.

\clearpage
\begin{figure}
\epsscale{}
 \plotone{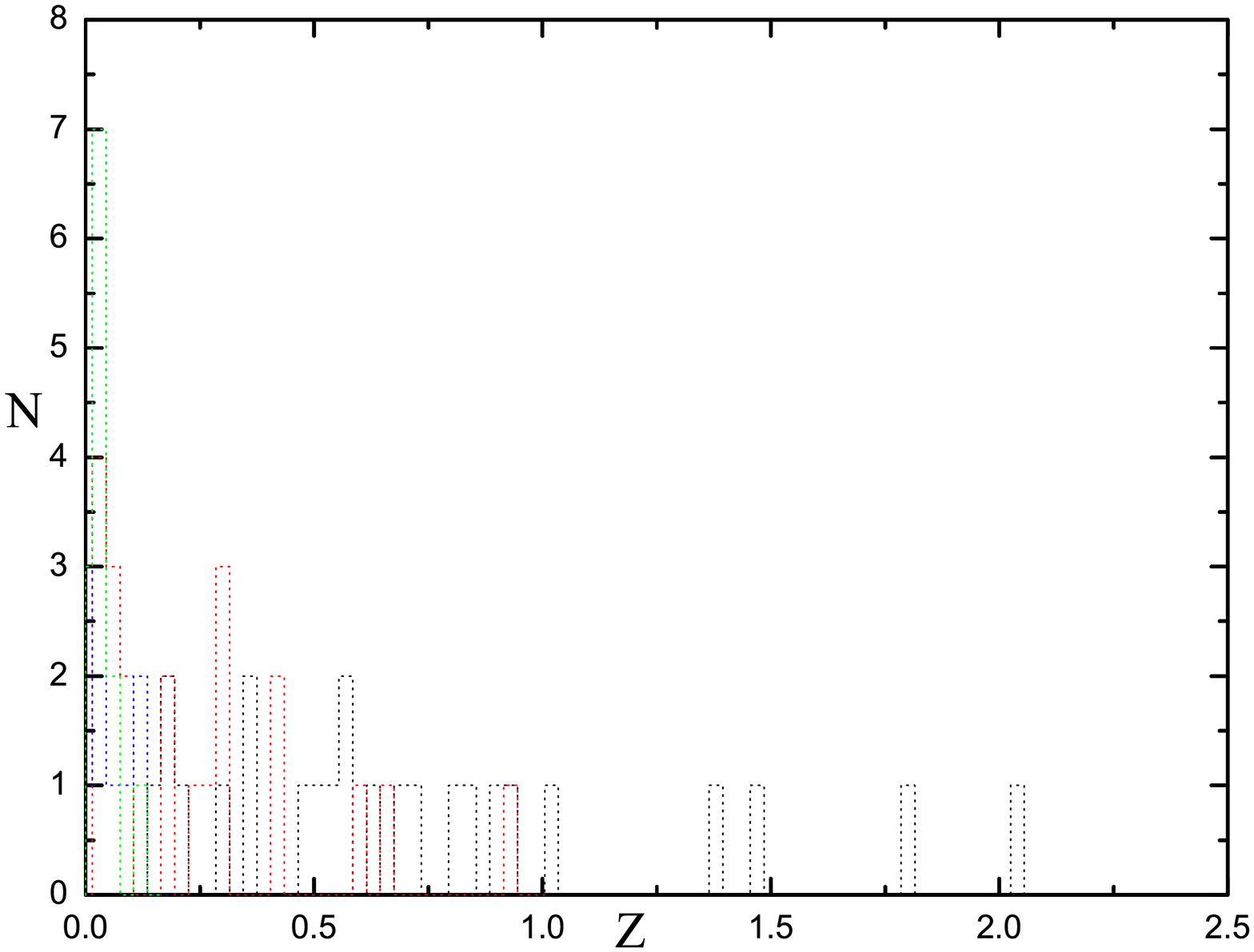}
 \caption{Redshift distributions for FSRQs (black dotted line),BL Lac objects (red dotted line),FRI (green dotted line) and FRII (blue dotted line)radio galaxies.}
\end{figure}

\begin{figure}
\epsscale{}
 \plotone{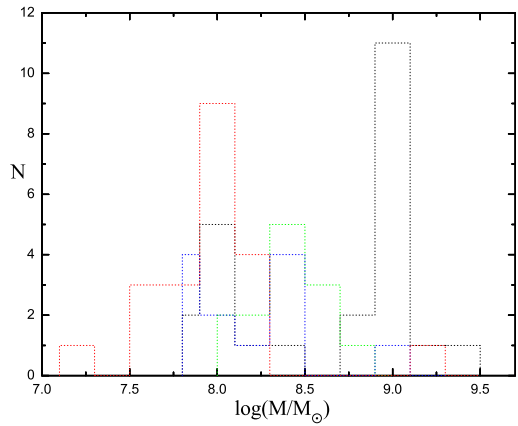}
 \caption{Black hole mass distributions for FSRQs, BL Lac objects, FRI and FRII radio galaxies. The meanings of different lines are as same Fig.1.}
\end{figure}

\begin{figure}
\epsscale{}
 \plotone{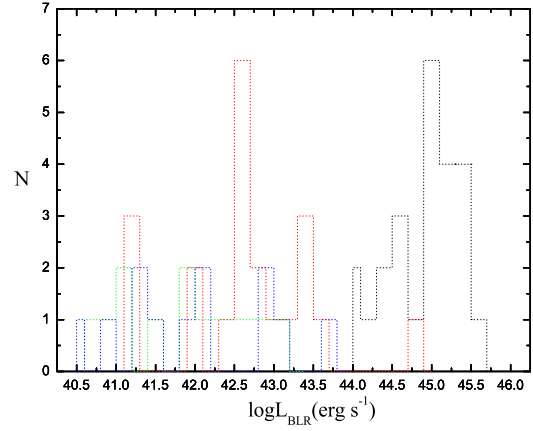}
 \caption{Broad line luminosity distributions for FSRQs, BL Lac objects, FRI and FRII radio galaxies. The meanings of different lines are as same Fig.1.}
\end{figure}

\begin{figure}
\epsscale{}
 \plotone{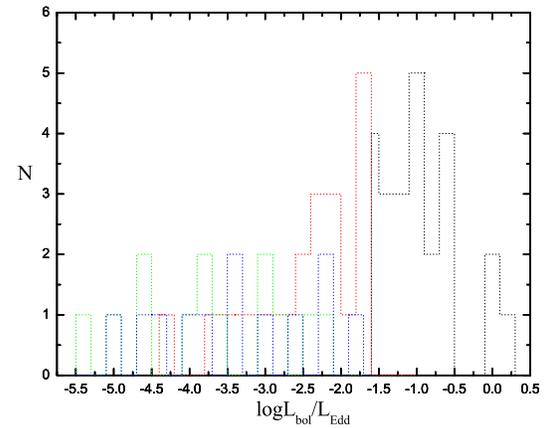}
 \caption{Eddington ratios distributions for FSRQs, BL Lac objects, FRI and FRII radio galaxies. The meanings of different lines are as same Fig.1.}
\end{figure}

\begin{figure}
\epsscale{}
 \plotone{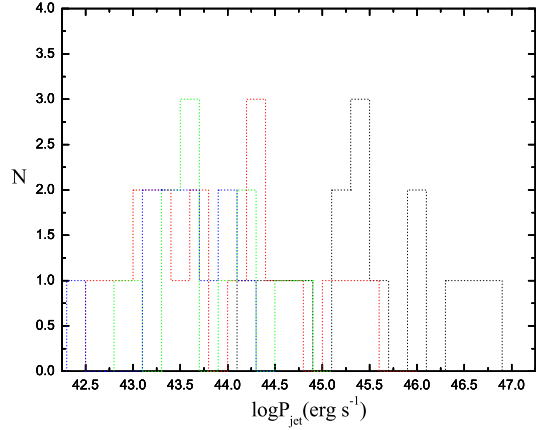}
 \caption{Jet power distributions for FSRQs, BL Lac objects, FRI and FRII radio galaxies. The meanings of different lines are as same Fig.1.}
\end{figure}

\begin{figure}
\epsscale{}
 \plotone{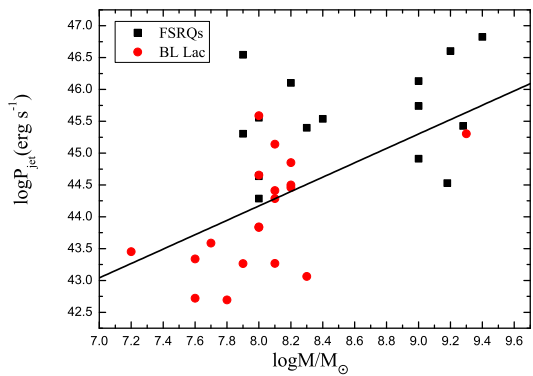}
 \caption{Correlation between the jet power and black hole mass for 24 FSRQs and 12 FRII radio galaxies. The black solid line represents the Ledlow-Owen dividing line between FRI and FRII radio galaxies given by Equation (4) with f=20.}
\end{figure}

\begin{figure}
\epsscale{}
 \plotone{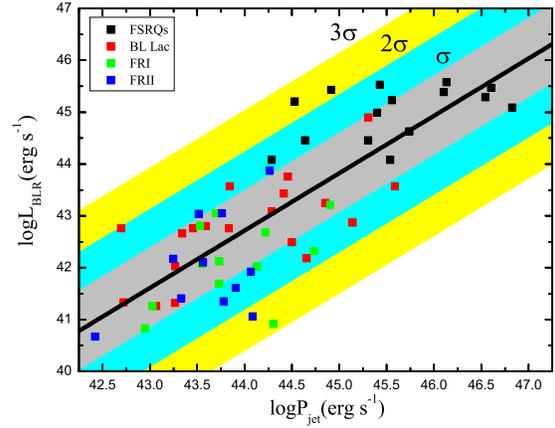}
 \caption{Correlation between the broad-line luminosity and jet power for 21 BL Lacs and 13 FRI radio galaxies. Shaded colored areas correspond to 1, 2 and 3$\rm{\sigma}$ (vertical) dispersion, $\rm{\sigma=0.77}$ dex. The black line is the best least square fit ($\rm{\log{L_{BLR}}}=\rm{1.10\log{P_{jet}}-5.997}$).}
\end{figure}

\clearpage
\begin{table*}
\small
\caption{The sample of blazars\label{tbl-1}}
\begin{tabular}{@{}crrrrrrrrrrr@{}}
\tableline
Name & Class & \rm{z} & $\log\Delta{t_{min}^{ob}}$ & Ref  & ${\rm{\log{L_{BLR}}}}$ & Ref & $\log{L_{core}^{5GHz}}$ & Ref &
$\rm{\log(M/{M_{\odot}})}$ & $\rm{\log{M_{\sigma}/M_{\odot}}}$\\
(1)& (2) & (3) & (4) & (5) & (6) & (7) & (8) & (9) & (10) & (11) \\
\tableline
0420-014	&	FSRQ	&	0.916	&	4.65	&	4	&	44.63	&	23	&	44.77 	&	28	&	9	&	9.03	\\
0528+134	&	FSRQ	&	2.06	&	4.94	&	5	&		&		&	46.46 	&	28	&	9.4	&		\\
0736+017	&	FSRQ	&	0.189	&	3.86	&	22	&	44.08	&	23	&	42.49 	&	28	&	8	&	8	\\
0827+243	&	FSRQ	&	0.9414	&	3.68	&	2	&	44.99	&	24	&	44.23 	&	28	&	8.3	&		\\
0846+513	&	FSRQ	&	0.5847	&	4.78	&	4	&		&		&		&		&	8.9	&		\\
0923+392	&	FSRQ	&	0.695	&	0.0698	&	22	&	45.53	&	25	&	44.28 	&	28	&	9.28	&	9.28	\\
1137+660	&	FSRQ	&	0.646	&	4.72	&	4	&	45.6	&	23	&		&		&	9	&	9.36	\\
1156+295	&	FSRQ	&	0.725	&	3.36	&	10	&	44.46	&	25	&	44.09 	&	28	&	7.9	&		\\
1226+023	&	FSRQ	&	0.158	&	4.57	&	4	&	45.43	&	23	&	43.48 	&	28	&	9	&		\\
1253-055	&	FSRQ	&	0.536	&	3.69	&	8	&	44.08	&	23	&	44.45 	&	28	&	8.4	&	8.43	\\
1355-416	&	FSRQ	&	0.313	&	4.78	&	4	&	45.04	&	23	&		&		&	9.15	&	9.73	\\
1406-074	&	FSRQ	&	1.494	&	4.76	&	12	&		&		&		&		&	9.2	&		\\
1510-089	&	FSRQ	&	0.36	&	3.5	&	13	&	44.46	&	23	&	43.04 	&	28	&	8	&	8.65	\\
1618+177	&	FSRQ	&	0.555	&	4.74	&	4	&		&		&		&		&	8.9	&		\\
1622-297	&	FSRQ	&	0.815	&	4.54	&	15	&		&		&		&		&	9.1	&		\\
1633+382	&	FSRQ	&	1.813	&	4.76	&	16	&	45.47	&	23	&	46.12 	&	28	&	9.2	&		\\
1641+399	&	FSRQ	&	0.593	&	3.65	&	4	&	45.23	&	23	&	44.48 	&	28	&	8	&		\\
1704+608	&	FSRQ	&	0.372	&	4.81	&	4	&	44.73	&	23	&		&		&	9	&		\\
2128-123	&	FSRQ	&	0.501	&	4.88	&	4	&	45.21	&	23	&	42.87 	&	28	&	9.18	&	9.61	\\
2135-147	&	FSRQ	&	0.2	&	4.86	&	4	&	44.65	&	23	&		&		&	9	&	8.94	\\
2141+175	&	FSRQ	&	0.211	&	3.69	&	4	&	44.26	&	24	&		&		&	8.1	&	8.74	\\
2223-052	&	FSRQ	&	1.404	&	3.48	&	4	&	45.29	&	23	&	46.02 	&	28	&	7.9	&		\\
2230+114	&	FSRQ	&	1.037	&	4.6	&	4	&	45.58	&	23	&	45.38 	&	28	&	9	&		\\
2251+158	&	FSRQ	&	0.859	&	3.84	&	21	&	45.39	&	23	&	45.34 	&	28	&	8.2	&	9.17	\\
0219+428	&	BL	&	0.444	&	3.4	&	1	&		&		&	43.07 	&	29	&	8	&		\\
0235+164	&	BL	&	0.94	&	3.4	&	2	&	43.57	&	26	&	44.53 	&	29	&	8	&		\\
0317+185	&	BL	&	0.19	&	3.75	&	3	&	42.6	&	26	&		&		&	8.1	&	8	\\
0548-322	&	BL	&	0.069	&	3.54	&	6	&		&		&	40.01 	&	29	&	7.8	&		\\
0716+714	&	BL	&	0.3	&	3.68	&	7	&	43.09	&	23	&	42.49 	&	28	&	8.1	&		\\
0735+178	&	BL	&	0.424	&	3.63	&	8	&	43.25	&	25	&	43.38 	&	28	&	8.2	&		\\
0754+100	&	BL	&	0.266	&	3.68	&	3	&	42.49	&	24	&	42.83 	&	28	&	8.2	&		\\
0851+202	&	BL	&	0.306	&	3.7	&	2	&	43.43	&	23	&	42.69 	&	29	&	8.1	&		\\
1101+384	&	BL	&	0.03	&	3.26	&	9	&	41.33	&	25	&	40.05 	&	29	&	7.6	&	8.29	\\
1219+285	&	BL	&	0.102	&	3.79	&	11	&	43.57	&	27	&	41.80 	&	29	&	8	&		\\
1235+632	&	BL	&	0.297	&	3.58	&	11	&	42.65	&	25	&		&		&	7.8	&		\\
1514-241	&	BL	&	0.049	&	2.95	&	4	&		&		&	41.19 	&	29	&	7.2	&	8.1	\\
1538+149	&	BL	&	0.605	&	3.86	&	14	&	42.87	&	23	&	43.83 	&	28	&	8.1	&	7.82	\\
1652+398	&	BL	&	0.034	&	4.1	&	8	&	41.26	&	25	&	40.59 	&	29	&	8.3	&	9.2	\\
1749+096	&	BL	&	0.322	&	3.69	&	1	&	43.76	&	26	&	42.76 	&	28	&	8.2	&		\\
1803+784	&	BL	&	0.68	&	4.94	&	3	&	44.9	&	26	&	44.09 	&	28	&	9.3	&		\\
1807+698	&	BL	&	0.051	&	3.38	&	17	&	41.32	&	25	&	40.90 	&	28	&	7.9	&	8.51	\\
2005-489	&	BL	&	0.071	&	3.84	&	18	&	42.03	&	26	&	40.90 	&	29	&	8.1	&	9.03	\\
2155-304	&	BL	&	0.116	&	2.95	&	19	&	42.66	&	25	&	41.02 	&	29	&	7.6	&		\\
2200+420	&	BL	&	0.069	&	3.44	&	20	&	42.8	&	26	&	41.40 	&	28	&	7.7	&	8.23	\\
2254+074	&	BL	&	0.19	&	3.68	&	14	&		&		&	41.79 	&	29	&	8	&	8.62	\\
\tableline
\end{tabular}
\tablecomments{(1) Xie et al. (1992); (2) Xie et al. (2001b); (3) Xie et al. (1991a); (4) Bassani, Dean, \& Sembay 1983; (5) Wagner et al. (1997); (6) Xie et al. (1996); (7) Qian,Tao,\& Fan . (2002); (8)Xie et al. (1999); (9) Xie et al. (1998); (10) Xie et al. (1994); (11) Xie et al. (1988); (12) Wagner et al. (1995); (13) Xie et al. (2002b); (14) Xie et al. (1990); (15) Mattox et al. (1997); (16) Fan et al. (1999); (17) Stauber,Brunner,\& Worrall. (1986); (18) Giommi et al. (1990); (19) Paltani et al. (1977); (20) Weistrop (1973); (21) Barbiberi, Romano, \& Zambon (1978); (22) Wu \& Urry (2002); (23) Cao \& Jiang (1999); (24) Sbarrato et al. (2012); (25) Celotti et al. (1997); (26) Wang et al. (2002); (27) Maraschi \& Tavecchio (2003); (28) Kharb et al. (2010); (29) Antonucci et al. (1985)}
\end{table*}

\clearpage
\begin{table*}
\small
\caption{The sample of radio galaxy\label{tbl-2}}
\begin{tabular}{@{}crrrrrrrrrrr@{}}
\tableline
Name & Class & \rm{z} & ${\rm{\log(M/{M_{\odot}})}}$ & Ref & ${\rm{\log{L_{BLR}}}}$ & Ref & $\log{L_{core}^{5GHz}}$ & Ref\\
(1)& (2) & (3) & (4) & (5) & (6) & (7) & (8) & (9)\\
\tableline
3C 78	&	FR I	&	0.029	&	8.5	&	2	&	42.02	&	3	&	40.25	&	4	\\
3C 84	&	FR I 	&	0.017	&	8.49	&	2	&	43.21	&	3	&	41.46	&	4	\\
3C 88	&	FR I	&	0.0301	&	8.03	&	2	&	43.05	&	3	&	39.57	&	4	\\
PKS 0453-206	&	FR I	&	0.035	&	8.55	&	2	&	41.26	&	3	&	38.53	&	3	\\
PKS 0915-11	&	FR I	&	0.054	&	8.69	&	2	&	42.12	&	3	&	39.63	&	3	\\
3C 29	&	FR I 	&	0.0451	&	8.2	&	2	&	41.69	&	3	&	39.63	&	3	\\
3C 89	&	FR I	&	0.139	&	8.52	&	2	&	42.68	&	3	&	40.39	&	4	\\
3C 120	&	FR I	&	0.033	&	8.13	&	1	&	42.32	&	3	&	41.19	&	3	\\
3C 338	&	FR I	&	0.03	&	8.78	&	2	&	42.08	&	3	&	39.34	&	4	\\
1333-331	&	FR I	&	0.013	&	8.77	&	2	&	40.92	&	3	&	40.52	&	3	\\
1610+296	&	FR I	&	0.032	&	8.96	&	2	&	40.83	&	3	&	38.4	&	3	\\
2236-176	&	FR I	&	0.074	&	8.49	&	2	&	41.18	&	3	&		&		\\
2322-122	&	FR I	&	0.082	&	8.33	&	2	&	42.81	&	3	&	39.31	&	3	\\
3C 33	&	FR II	&	0.06	&	8.38	&	2	&	42.1	&	3	&	39.36	&	4	\\
3C 98	&	FR II	&	0.031	&	7.88	&	2	&	42.17	&	3	&	38.87	&	4	\\
3C 223	&	FR II	&	0.137	&	8.15	&	2	&	41.35	&	3	&	39.7	&	4	\\
3C 293	&	FR II	&	0.045034	&	7.99	&	2	&	43.05	&	3	&	39.67	&	4	\\
PKS 0131-360	&	FR II	&	0.03	&	8.53	&	2	&	41.06	&	3	&	40.18	&	3	\\
3C 388	&	FR II	&	0.091	&	9.18	&	2	&	41.92	&	3	&	40.15	&	4	\\
3C 390.3	&	FR II	&	0.0561	&	8.55	&	2	&	43.87	&	3	&	40.46	&	4	\\
PKS 0442-28	&	FR II	&	0.147	&	8.48	&	2	&	43.11	&	3	&		&		\\
PKS 0449-175	&	FR II	&	0.031	&	8.48	&	2	&	40.67	&	3	&	37.58	&	3	\\
0123-01	&	FR II	&	0.018	&	7.86	&	2	&	41.61	&	3	&	39.9	&	3	\\
0634-206	&	FR II	&	0.056	&	8.09	&	2	&	43.04	&	3	&	39.29	&	3	\\
1448+63	&	FR II	&	0.042	&	7.92	&	2	&	41.41	&	3	&	39	&	3	\\
\tableline
\end{tabular}
\tablecomments{(1) Xie et al. (2004b);(2) Woo \& Urry (2002); (3) Zirbel \& Baum (1995); (4)Buttiglione et al. (2010);}
\end{table*}
\end{document}